\documentclass[superscriptaddress,amsmath,amssymb,nofootinbib,twocolumn,floatfix]{revtex4-1}

\usepackage{setspace}
\usepackage{graphicx}  
\usepackage{dcolumn}   
\usepackage{bm}        
\usepackage{rotating}
\usepackage{lipsum}

\newcommand{\sectionname}[1]{ \noindent{\bfseries #1}.---}

\begin{document}


\title{Measurement of the Stray Light in the Advanced Virgo Input Mode Cleaner Cavity using an instrumented baffle}

\author{O. Ballester}
\author{O. Blanch}
\author{L. Cardiel}
\author{M. Cavalli-Sforza}
\affiliation{Institut de F\'\i sica  d'Altes Energies (IFAE), Barcelona Institute of Science and Technology, E-08193 Barcelona, Spain}
\author{A. Chiummo}
\affiliation{European Gravitational Observatory (EGO), Cascina, Pisa, Italy}
\author{C. Garc\'\i a}
\author{J.M. Illa }
\author{C. Karathanasis}
\author{M. Kolstein}
\affiliation{Institut de F\'\i sica  d'Altes Energies (IFAE), Barcelona Institute of Science and Technology, E-08193 Barcelona, Spain}
\author{M. Mart\'\i nez}
\affiliation{Institut de F\'\i sica  d'Altes Energies (IFAE), Barcelona Institute of Science and Technology, E-08193 Barcelona, Spain}
\affiliation{Instituci\'o Catalana de Recerca i Estudis Avançats (ICREA), Barcelona, Spain}
\author{A. Men\'endez-V\'azquez}
\author{Ll. M. Mir}
\author{J. Mundet}
\author{A. Romero-Rodr\'\i guez}
\author{D. Serrano}
\affiliation{Institut de F\'\i sica  d'Altes Energies (IFAE), Barcelona Institute of Science and Technology, E-08193 Barcelona, Spain}
\author{H. Yamamoto}
\affiliation{LIGO laboratory, California Institute of Technology (Caltech), Pasadena, CA, US}
\
\date{\today}


\begin{abstract}

A new instrumented baffle was installed in Spring 2021 at Virgo surrounding the suspended mirror in the input mode cleaner triangular cavity.  
It serves as a demonstrator of the technology designed to instrument the baffles in the main arms in the near future. 
We present, for the first time, results on the measured scattered light distribution inside the cavity as determined by 
the new device using data collected between May and July 2021, with Virgo in commissioning phase and operating with an 
input laser power in the cavity of 28.5~W. The sensitivity of the baffle is discussed and the data is compared to scattered light simulations.

\end{abstract}


\maketitle

\begin{figure}[htb]
\begin{center}
\mbox{
 \includegraphics[width=0.23\textwidth]{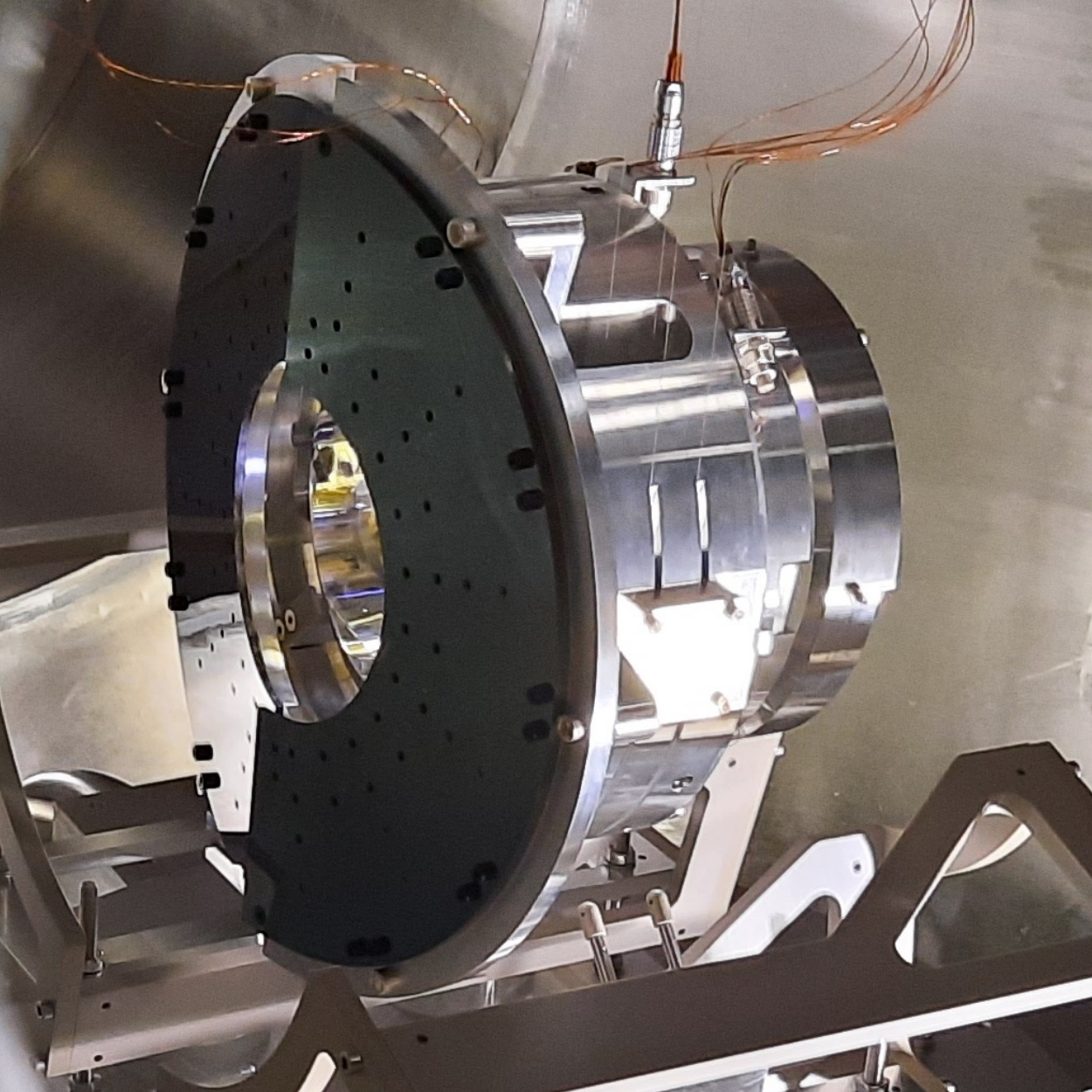}
 \includegraphics[width=0.23\textwidth]{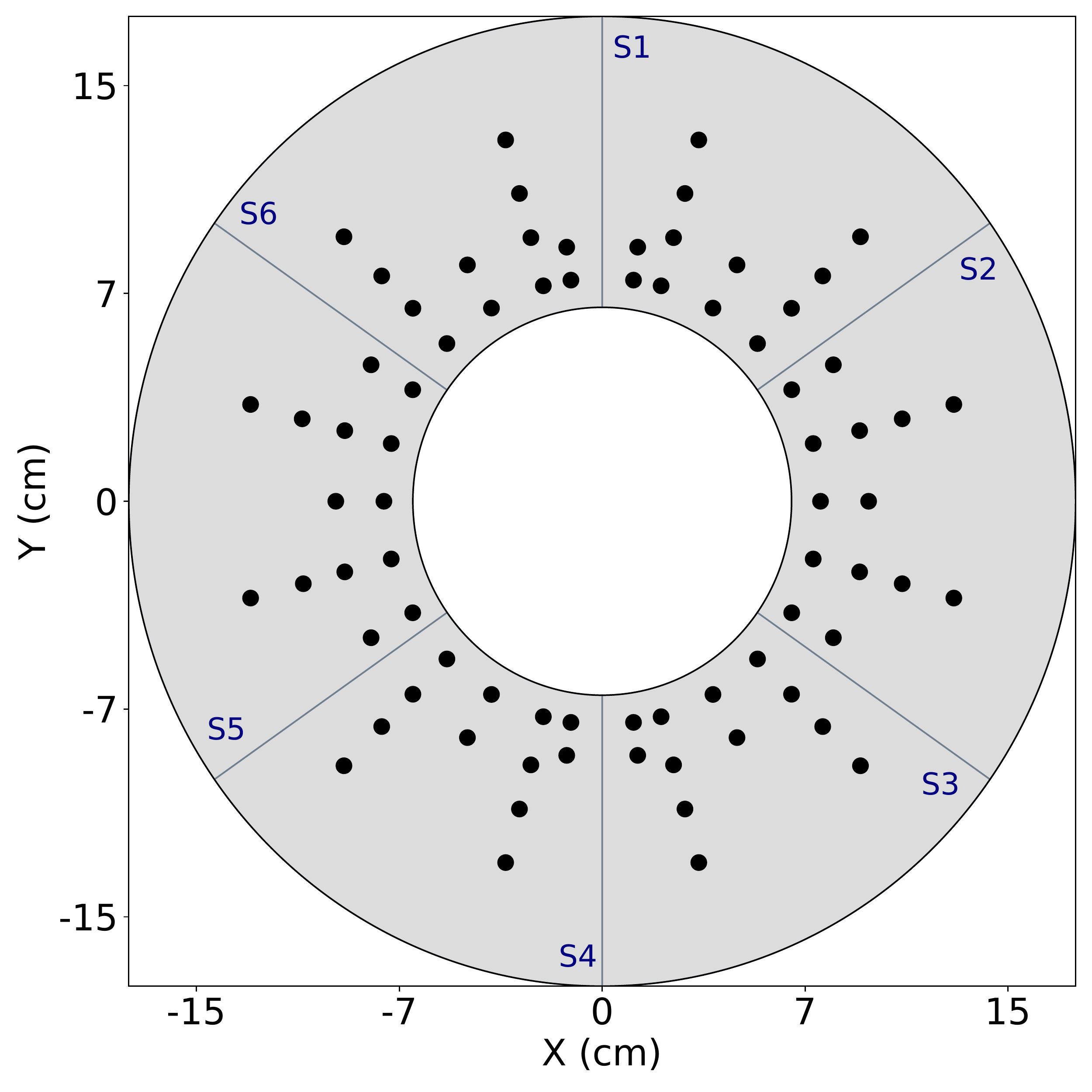}
}
\end{center}
\caption{\small
(left) Picture of the instrumented baffle installed inside the IMC end mirror vacuum tower.
(right) Sketch  of  the location of the photosensors in the baffle in the $x-y$ plane. 
The gray lines define the six regions where sensors signals are added to determine various observables in the data and simulation. 
}
\label{fig:imc}
\end{figure}

\sectionname{Introduction} 
\label{sec:intro}
As part of the phase II upgrade of the Advanced Virgo interferometer~\cite{TheVirgo:2014hva,Virgo:2019juy}, the experiment plans to equip the suspended areas 
surrounding the main test masses in the Fabry-Perot cavities  with instrumented baffles.  This opens up the possibility of monitoring the scattered 
light from the mirrors in the cavity at low angles, providing a dynamic mapping of the mirror surface and defects. 
In addition, the baffles would assist in the alignment of the cavities, in the detection of higher-order laser modes, and in establishing 
correlations with interferometer glitches. 
In April 2021, a first instrumented baffle was installed surrounding the suspended end mirror of the Virgo's input mode cleaner (IMC) cavity
(see Figure~\ref{fig:imc}). This intervention took place in the framework of phase I upgrades in the IMC, including a new suspension 
payload and a new end mirror.  This new baffle serves as a demonstrator of the technology for instrumenting the baffles in Virgo's main arms.
This paper presents first results on the performance of the detector and the comparison with the simulations describing the scattered light 
inside the IMC cavity.

\sectionname{Instrumented Baffle}
\label{sec:instrument}
A very detailed description of the baffle technology will be  provided in a separate publication. 
Here a brief description is given. 
The baffle has an inner radius (in the $x-y$ plane) of 7~cm and an outer radius of 17.5~cm. It is divided into 
two halves, each tilted 9~degrees with respect to the normal to the nominal direction of the laser beam to avoid complete back reflections in the cavity. 
The infrared light from the IMC cavity penetrates the holes in the mirror-polished stainless steel baffle reaching Si-based $7.37 \times 7.37 $~mm${}^2$
(sensitive area $6.97 \times 6.97$~mm${}^2$) newly developed photosensors,  provided by  the Hamamatsu company in Japan. 
The holes have a conical shape centered around the sensors with 4~mm diameter at the front baffle surface, such that the light 
does not resolve the sensor edge and the scattering of light with the hole geometry is minimized.  Both the  baffle and the sensor 
surfaces include  anti-reflecting coatings at 1064~nm. The baffle is equipped with 76~sensors symmetrically placed with respect to 
the $y$ axis and  mounted on two large gold-plated polyamide-based PCBs. All the elements in the baffle are certified for ultra-high 
vacuum (UHV) conditions. 

The layout of the sensors in the $x-y$ plane is presented in Figure~\ref{fig:imc}.  The sensors are mostly located in two concentric rings at radii 
of 8.8 and 9.8~cm, respectively.  Additional sensors are distributed across the baffle at larger radii and different azimuthal angles.  
The sensor signals are processed by 16~ADCs (8~ADCs in each half baffle), which at present average each of the 
sensor signals over 500~ms, translating into a baffle readout rate of 2~Hz, with the potential to reach up to 10~Hz in the future.  
The calibration of the photosensors in the laboratory,  using a dedicated laser setup, 
indicated a good linearity in the response for the whole range of interest and a less than 3$\%$ sensor-to-sensor variation. 
The absolute calibration was determined to be $4.6~\mu$W/ADC~count, 
with an uncertainty of about 5$\%$. 
In addition, each ADC is instrumented with a temperature sensor, leading to 16 separate readings. 
The whole system operates with a limited voltage of 3.3~V.  

The performance of the IMC cavity was checked right after the instrumented baffle was installed and 
the operations of the IMC cavity restored. The level of losses and the throughput power in the cavity showed no 
significant changes.
Currently, the instrumented baffle is an integral part of the Virgo operations. 

\begin{figure}[htb]
\begin{center}
\mbox{\includegraphics[width=0.47\textwidth]{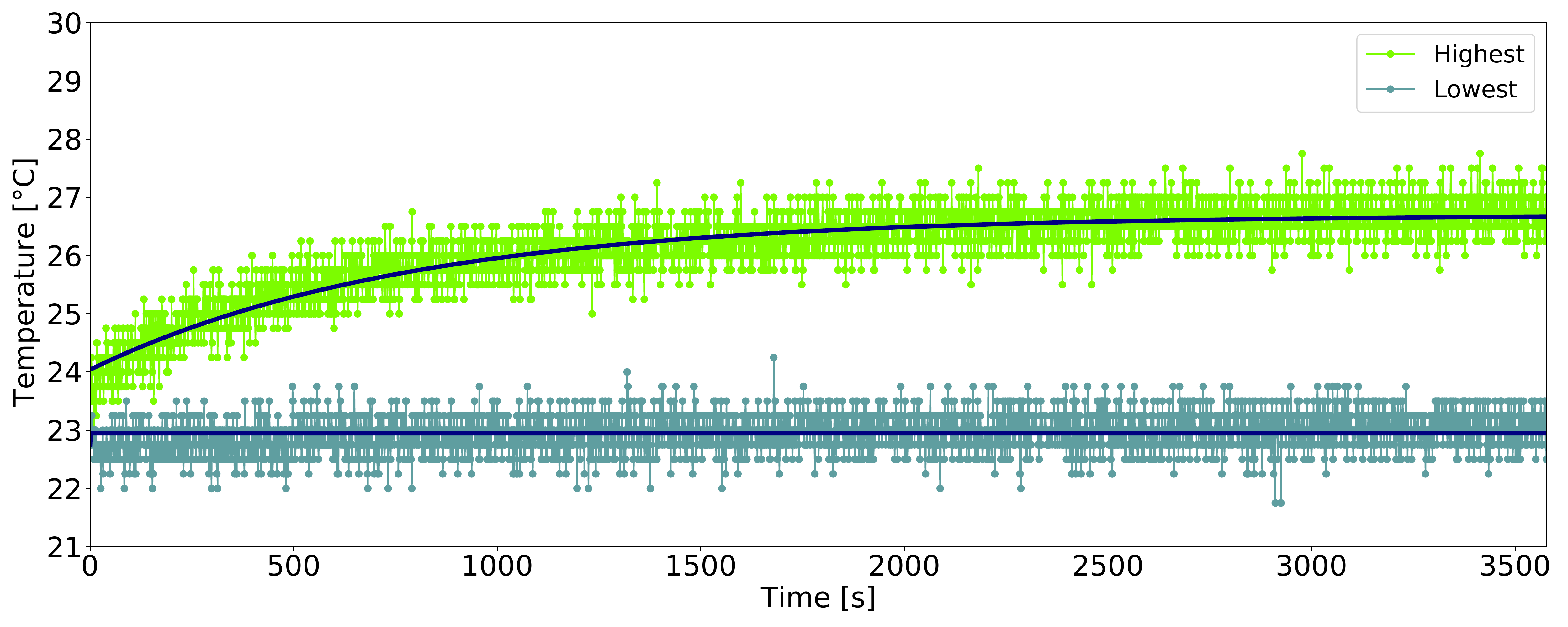}
}
\end{center}
\caption{\small
Measurement of the temperature as a function of time for a period of one hour.
The ADCs showing the highest and lowest temperatures are shown.
}
\label{fig:temp}
\end{figure}

The suspended baffle operates at room temperature and under UHV conditions, with no active cooling possible, given the limitations dictated by the suspension. 
Therefore, the potential overheat of the front-end electronics is a priori an important aspect. Thanks to the moderate operating voltage, 
the efficient heat dissipation of the gold-plated PCB, and a careful design of 
the mechanical couplings with the stainless steal structure, acting as cold mass, no overheating is observed. 
This is illustrated in Figure~\ref{fig:temp}, 
where the readings of two temperature sensors are displayed. The system reaches thermal equilibrium after 40 minutes with temperatures in the range between 22 and 28°C. 
The sensor with the highest readings corresponds to that closest to the micro-controller electronics, 
dissipating most of the heat.  

\begin{figure}[htb]
\begin{center}
\mbox{
\includegraphics[width=0.22\textwidth]{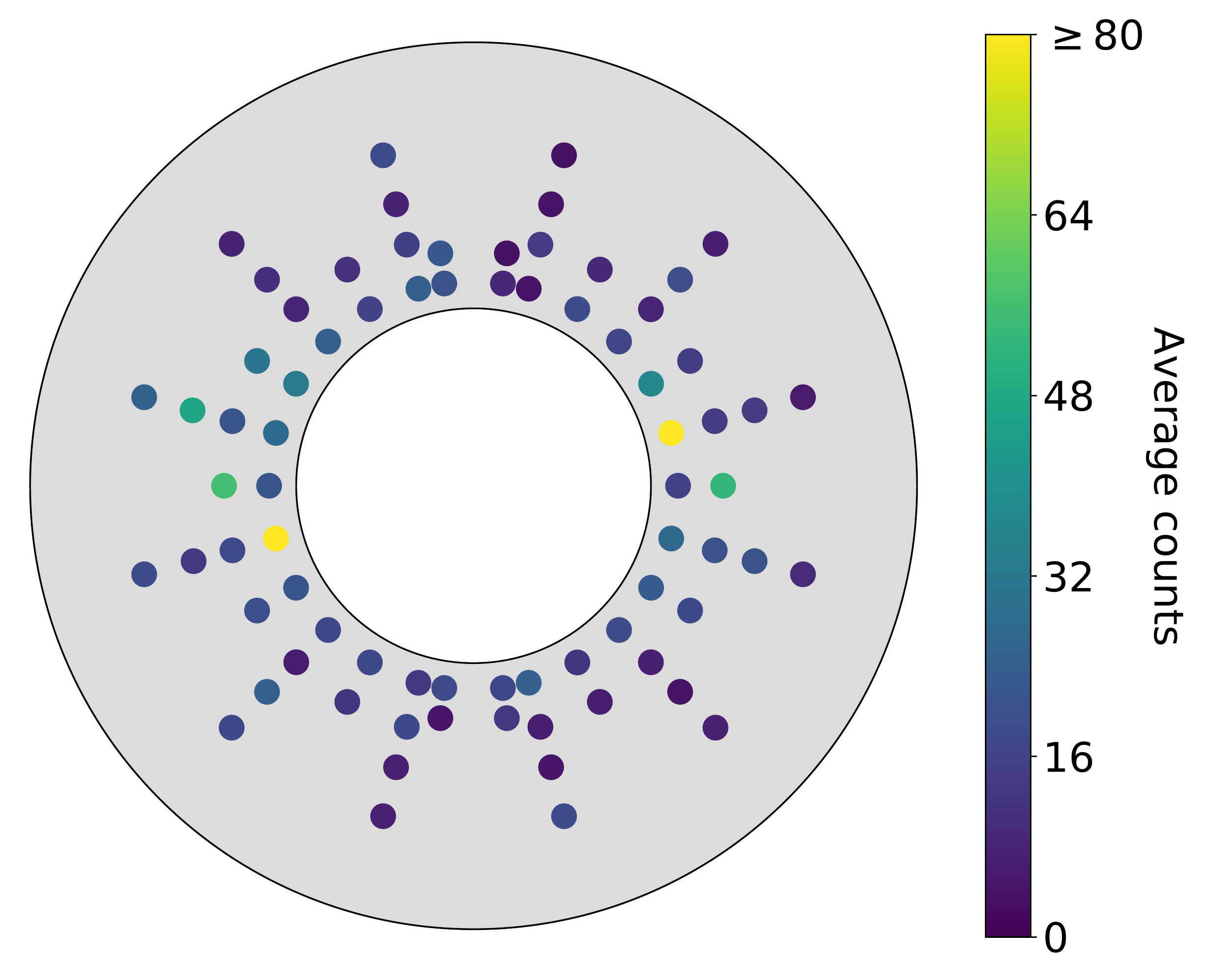}
\includegraphics[width=0.22\textwidth]{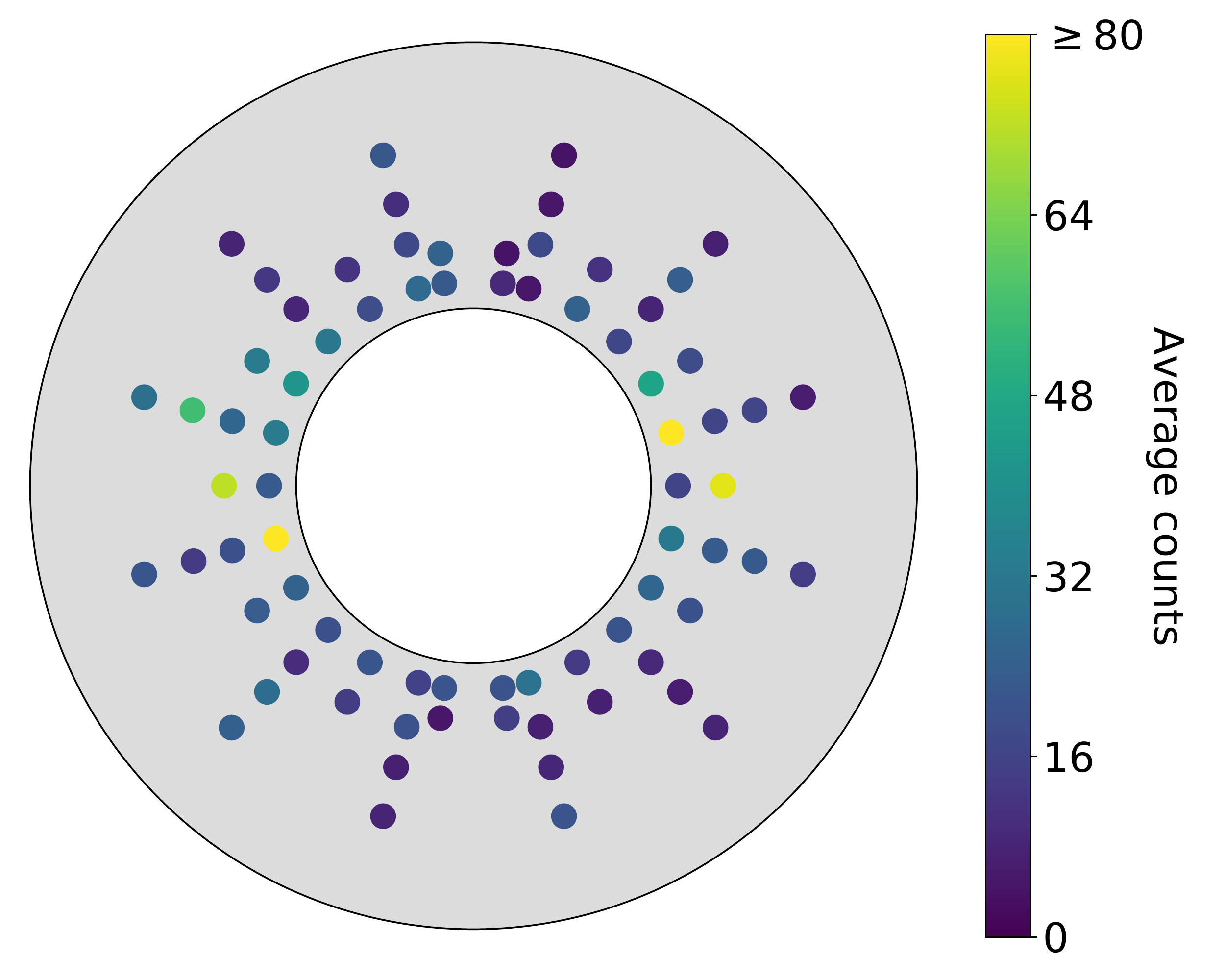}
}
\end{center}
\caption{\small
Observed light power in the individual sensors, in ADC counts, averaged over time, for two separate data sets spanning several weeks. 
}
\label{fig:sensors}
\end{figure}

\sectionname{Raw Measurements}
\label{sec:results}
The baffle showed good performance in the absence of light in the cavity with average noise levels in the channels limited to up to seven counts, 
with an RMS of 0.01 to 0.16 counts, depending on the sensor, 
and a signal-to-noise ratio typically in the range between one and more than ten in the presence of light. 
The stability of the baffle signals with time was studied indicating that they remain constant over time. 
The raw signals in the individual photosensors in ADC counts, 
averaged over the period of one hour and with no noise-suppression applied, 
are presented in Figure~\ref{fig:sensors} for two separate data sets. As expected,  the 
signals are concentrated at low radius with sensors reaching more than 100~counts. 
The data show a left-right asymmetry, with more power in the half baffle at negative $x$ values. 
In addition, the data suggest the power is concentrated in a plane tilted by about 15~degrees in the $\phi$ direction with respect to the nominal 
$x-z$ plane of the triangular cavity.  These effects are persistent in all the data analysed.  

Few special runs were taken after forcing small displacements (of the order of few $\mu$m) of the IMC end mirror with respect to its nominal position 
while maintaining the cavity at resonance. This translated into changes in the IMC output power and the appearance of laser higher-order modes, which 
in turn resulted into about 5$\%$ variations in the total power measured in the baffle, thus showing the sensitivity of the baffle to the status of the cavity. 
Similarly, the baffle proved to be quite sensitive to momentary unlocks that lead to the absence of circulating light in the cavity, resulting 
in an abrupt decrease in baffle signals.

\sectionname{Comparison with Simulations}
\label{sec:sim}
The average noise pedestals in each of the channels, as determined during quiet periods with no laser in the IMC cavity, are subtracted channel-by-channel 
in the data before the final absolute calibration is applied. 
An additional noise-suppression algorithm is a priori applied with the aim to remove 
the channels with signals compatible with noise within two standard deviations of the noise. 
However, no channel is removed from the data because of an excellent 
signal-to-noise performance of the instrument. 

\begin{figure}[htb]
\begin{center}
\mbox{
 \includegraphics[width=0.22\textwidth]{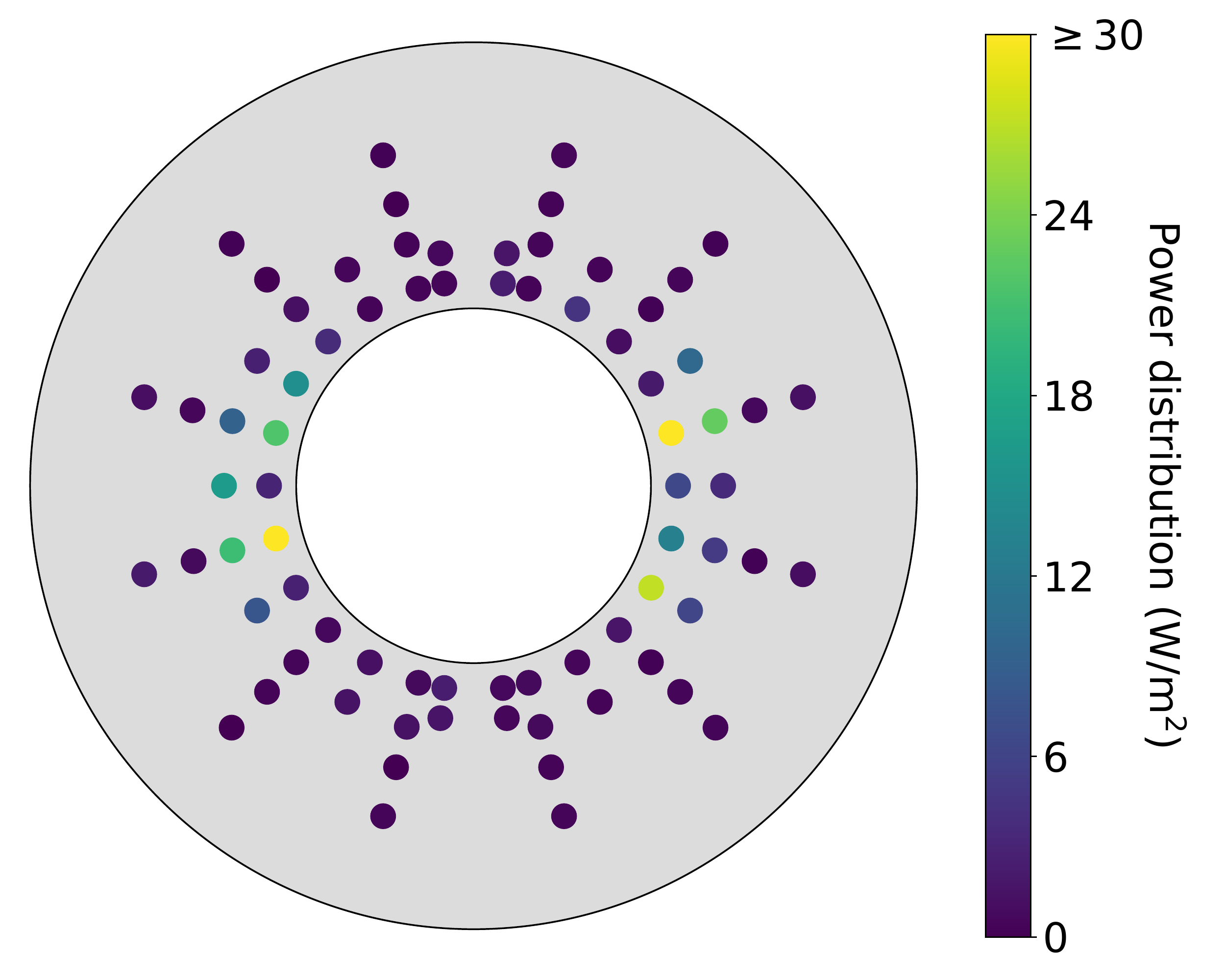}
}
\end{center}
\caption{\small
 2-D map of the simulated power distribution (in W/m$^2$) in the baffle surface.  
}
\label{fig:sim_baffle}
\end{figure}

The calibrated data are compared to simulations~\cite{Yamamoto_2020} of the light inside the IMC cavity.  
The simulation follows closely that presented in Ref.~\cite{Romero:2020ovg}, with updated mirror maps 
and considering a nominal input laser power of $28.5\pm 0.1$~W, corresponding to the average input power in the cavity determined during the data 
taking period under consideration with an uncertainty of $0.2\%$.   
The latter is included in the calibrated data as an additional uncertainty.
The simulation does not include thermal effects in the mirrors induced by the laser,  that are expected to be small. 
As discussed in Ref.~\cite{Romero:2020ovg}, 
the baffle itself is only approximately simulated as a geometrical position in which the intensity of the field is evaluated. 
Figure~\ref{fig:sim_baffle} shows the 2-D distribution of light (in $\mathrm{W/m^2}$) in the baffle area. 
The power is concentrated along the $x$-axis (see also Table~\ref{tab:table} below). 
In order to determine a possible dependency of the simulated fields with the end mirror maps, 
the results are compared with those from Ref.~\cite{Romero:2020ovg}, 
including previous mirror maps, resulting into a negligible difference. 

\begin{figure}[htb]
\begin{center}
\mbox{
\includegraphics[width=0.47\textwidth]{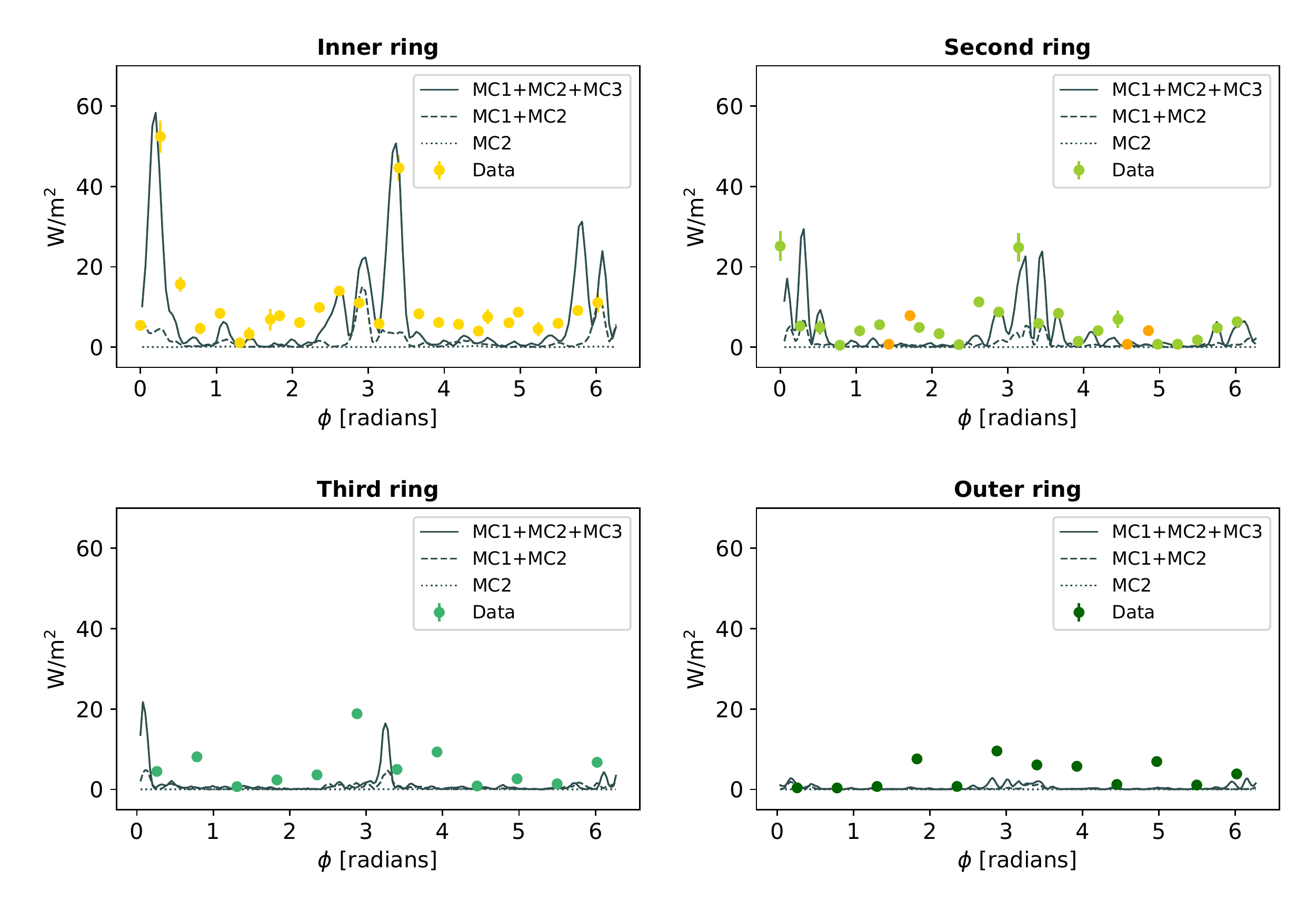}
}
\end{center}
\caption{\small
Measured differential distribution of the power intensity in the baffle 
(dots) as a function of $\phi$ for four different radii. 
The second ring includes four sensors at a slightly different radius 
(see Figure~\ref{fig:imc}) at $\phi \sim \pi/2$ and $\phi \sim 3\pi/2$.
Data are compared to simulations (gray lines) obtained 
including the three mirrors (MC1+MC2+MC3) in the IMC cavity,
removing the dihedron right mirror (MC1+MC2) or
only including the end mirror (MC2) surface maps.
}
\label{fig:dphi}
\end{figure}

Figure~\ref{fig:dphi} shows the measured differential distribution of the power intensity in the baffle as a function of 
$r$ and $\phi$ compared to simulations, where $r$ and $\phi$ are evaluated at the center of each sensor, and only the active area of the sensor 
($\pi \times 4$~mm${}^2$) is taken into account.  The data present a strong dependence with $r$,  
with the power being concentrated in the inner part. 
The power intensity observed varies in the range between 
1.1~$\mathrm{W/m^2}$ and 52.3~$\mathrm{W/m^2}$ at small $r$,  
and between 0.4~$\mathrm{W/m^2}$ and 9.6~$\mathrm{W/m^2}$ at very large $r$. 
As already anticipated, the data in the two innermost rings present a tilted $\phi$ distribution with 
two prominent structures in the vicinity of $\phi \sim 0$ and $\phi \sim \pi$. 

The simulation, including the information of the mirror surface maps for all the mirrors (M1, M2, M3) in the IMC triangular cavity, 
provides a fair description of the main features in the data but fails to describe the
total power in the baffle.  Table~\ref{tab:table} presents the results integrated over the sensors in each sextant.
In the case of the simulation, the results are only presented normalised  by  the power  circulating  in  the  cavity (9695 W), which factorizes out uncertainties.
A total power in the sensors of about 6.6~mW is observed, whereas the simulation predicts around 4.4~mW. 
As shown in Table~\ref{tab:table}, the predictions are close to the data in sextants S2 and S5, corresponding to those with the 
largest observed power, but underestimates the data in the rest of sextants. 
Similarly, the simulation does not completely support the level of left-right asymmetry observed in the data by 
comparing the information in the different sextants.  

The measured tilt in $\phi$ cannot be attributed to  a misalignment of the triangular cavity in data 
and is presumably dominated by the details of the mirror surface maps.  
In particular,  
the dihedron mirrors (here denoted as M1 and M3) 
are located  143~m away from the end mirror (denoted as M2) in the IMC cavity, 
and are facing the instrumented baffle.  
For illustration purposes, 
the simulations were repeated assuming perfect dihedron mirrors and therefore ignoring their mirror surface map information. 
As shown in Figure~\ref{fig:dphi}, 
this translates into the disappearance of any $\phi$ dependence in the simulated power intensity. 
Therefore, 
we conclude that the disagreement between data and simulation is most probably originated by 
a non-accurate enough description of the mirror surface maps in the dihedron,
mainly that of the MC3 mirror,
which is responsible for the larger energy deposits in some of the PDs.
Thus, 
this comparison will help establish the requirements on future IMC mirror surface specifications.
At large radius,  
the amount of light detected is small and the predictions might be also affected by large angle scattering contributions not included in the simulation. 
Other important dynamics,
such as polarisation mixing might also be missing,
and some stray light underestimated.
More data will indicate whether additional work is required in the procedure to subtract detector pedestals.
These are all important topics that will be studied using the data measured by this instrumented baffle.

\begin{table}[ht]
\begin{center}
\begin{small}
\begin{tabular}{c c c c} 
\hline
Sextant & \multicolumn{2}{c}{Data}  & Simulation\\ 
\hline
   & Power (mW) &   Ratio ($\times 10^{-5}$) &  Ratio ($\times 10^{-5}$)\\ 
\hline
S1 &  $0.47 \pm 0.02$  &   $4.8 \pm 0.2$  &   1.0  \\
S2 &  $1.92 \pm 0.10$  &  $19.8 \pm 1.0$  &  19.6  \\
S3 &  $0.55 \pm 0.03$  &   $5.7 \pm 0.3$  &   1.6  \\
S4 &  $0.67 \pm 0.03$  &   $6.9 \pm 0.4$  &   1.1  \\
S5 &  $2.25 \pm 0.11$  &  $23.2 \pm 1.2$  &  20.6  \\
S6 &  $0.76 \pm 0.04$  &   $7.8 \pm 0.4$  &   1.5  \\ 
\hline
\end{tabular}
\end{small}
\caption{\small
Measured power distribution in the different baffle sextants (see Figure~\ref{fig:imc}) compared to simulations. 
The second column shows the total power, whereas the rightmost columns (``Ratio'') show the total power normalised by the power circulating
in the cavity (9695 W).
The uncertainty on the data is dominated by the uncertainty in the calibration of the sensors (5.0\%) as the uncertainty on the IMC input power (0.2\%) is negligible.
}
\label{tab:table}
\end{center}
\end{table}

\sectionname{Conclusions}
\label{sec:conclu}
We have presented first results from the new instrumented baffle installed in the Virgo's input mode cleaner end mirror in 2021.
The data collected by the baffle have been used to measure, 
for the first time, 
the scattered light inside the cavity.  
The measured light distribution in the baffle presents features dominated by 
scattering processes from the mirrors facing the
instrumented baffle in the cavity. 
The data have been compared to simulations that provide an approximate description of the data.  
These results will serve to calibrate the simulations and demonstrate already the potential of 
instrumented baffles to detect defects in the mirrors and to improve the understanding of the scattered light inside ground-based gravitational wave 
experiments like Virgo.

\sectionname{Acknowledgements}
The  authors  gratefully  acknowledge  the  European  Gravitational  Observatory(EGO) and the Virgo Collaboration for providing access to the facilities and the support from EGO and the University of Pisa during the installation of the new instrument in Virgo. The LIGO  Observatories  were  constructed  by  the  California Institute  of  Technology and Massachusetts Institute of Technology with funding from the National Science  Foundation  under  cooperative  agreement  PHY-9210038.   The  LIGO Laboratory  operates  under  cooperative  agreement  PHY-1764464.
This work is partially  supported   by  the Spanish MINECO   under
the grants SEV-2016-0588, PGC2018-101858-B-I00,  and PID2020-113701GB-I00 some of which include
ERDF  funds  from  the  European  Union. IFAE  is  partially funded by
the CERCA program of the Generalitat de Catalunya. 


\bibliographystyle{apsrev}
\bibliography{baffle}{}

\end{document}